\documentclass[journal, english]{IEEEtran}
\usepackage[LGR,T1]{fontenc}
\usepackage[latin9]{inputenc}
\usepackage{array}
\usepackage{amsmath}
\usepackage{amssymb}
\usepackage{mathdots}
\usepackage{graphicx}
\usepackage{esint}
\PassOptionsToPackage{version=3}{mhchem}
\usepackage{mhchem}
\usepackage{supertabular,booktabs}
\usepackage{gensymb}

\makeatletter

\usepackage{subfigure}
\usepackage{color}
\providecommand{\tabularnewline}{\\}

\usepackage{todonotes}
\makeatother

\usepackage{babel}

\begin{document}

\author{Robert~R.~Richardson
	and~David~A.~Howey,~\IEEEmembership{Member,~IEEE}% <-this % stops a space
	\thanks{The authors are with the Department
		of Engineering Science, University of Oxford, Oxford, 
		UK, e-mail: \{robert.richardson, david.howey\} @eng.ox.ac.uk.
		
		Copyright \copyright \hspace{0.01 cm} 2015 IEEE}}%
	%\thanks{Manuscript received ; revised }}

%% The paper headers
%\markboth{Journal of \LaTeX\ Class Files,~Vol.~11, No.~4, December~2012}%
%{Shell \MakeLowercase{\textit{et al.}}: Bare Demo of IEEEtran.cls for Journals}

\title{Sensorless Battery Internal Temperature Estimation using a Kalman Filter with
Impedance Measurement}

\maketitle

\begin{abstract}
This study presents a method of estimating battery cell core and surface temperature using a thermal model coupled with electrical impedance measurement, rather than using direct surface temperature measurements. This is advantageous over previous methods of estimating temperature from impedance, which only estimate the average internal temperature. The performance of the method is demonstrated experimentally on a 2.3 Ah lithium-ion iron phosphate cell fitted with surface and core thermocouples for validation. An extended Kalman filter, consisting of a reduced order thermal model coupled with current, voltage and impedance measurements, is shown to accurately predict core and surface temperatures for a current excitation profile
based on a vehicle drive cycle. A dual extended Kalman filter (DEKF) based on the same thermal model and impedance measurement input is capable of estimating the convection coefficient at the cell surface when the latter is unknown. The performance of the DEKF using impedance as the measurement input is comparable to an equivalent dual Kalman filter using a conventional surface
temperature sensor as measurement input.\end{abstract}

\section{Introduction\label{sec:Introduction}}
\IEEEPARstart{T}{he} sustainable development of transportation relies on the widespread adoption of electric vehicle (EV) and hybrid electric vehicle (HEV) technology. Lithium-ion batteries are suitable for these applications due to their high specific energy and power density. However, their widespread deployment requires reliable on-board battery management systems to ensure  safe and optimal performance. In particular, accurate on-board estimation of battery temperature is of critical importance.
Under typical operating conditions, such as a standard vehicle drive cycle,
cells may experience temperature differences between surface and core
of 20 $^{\circ}$C or more \cite{Kim2009}. High battery temperatures
could trigger thermal runaway resulting in fires,
venting and electrolyte leakage. While such incidents are rare \cite{Wang2012a},
consequences include costly recalls and potential endangerment of
human life.
%It is therefore of great importance to have a reliable
%on-board thermal management system with accurate cell temperature
%estimation.

The conventional approach to temperature estimation is to use numerical
electrical-thermal models \cite{Forgez2010a,Kim2014b,Lin2013f,Kim2013,Lin2014}.
Such models rely on knowledge of the cell thermal properties, heat
generation rates and thermal boundary conditions. Models without online sensor
feedback are unlikely to work in practice since their temperature
predictions may drift from the true values due to small uncertainties
in measurements and parameters. However, using additional online measurements - typically of the cell surface temperature and of the temperature
of the cooling fluid - coupled with state estimation techniques such
as Kalman filtering, the cell internal temperature may be estimated
with high accuracy \cite{Kim2014b,Lin2013f,Kim2013,Lin2014}. However,
large battery packs may contain several thousand cells \cite{Pesaran2009},
and so the requirement for surface temperature sensors on every cell
represents substantial instrumentation cost.

An alternative approach to temperature estimation uses electrochemical
impedance spectroscopy (EIS) measurements at one or several frequencies to
directly infer the internal cell temperature, without using a thermal
model \cite{Srinivasan2012a,Schmidt2013a,Raijmakers2014d,Srinivasan2011c,Zhu2015}. This exploits the fact that impedance is related to a type of volume averaged cell temperature, which we define later in this article. 
For brevity, we refer to the use of impedance to infer such a volume-averaged temperature as `Impedance-Temperature Detection' (ITD). This has promise for practical application, since methods capable of measuring EIS spectra using
existing power electronics in a vehicle or other application have been developed
\cite{Howey2014a,Brandon2012,Huang2014c}. However, just like conventional
surface temperature sensors, ITD alone does not provide a unique solution
for the temperature distribution within the cell. Our previous work
showed that by combining ITD with surface temperature measurements
the internal temperature distribution could be estimated \cite{Richardson2014}.
However, this approach still requires each cell to be fitted with a surface temperature sensor.
Moreover, whilst the ITD technique was validated under
constant coolant temperature conditions, the accuracy of the technique
may be reduced if the temperature of the cooling medium is varied
rapidly, as discussed in Section \ref{sec:Frequency-Domain-Analysis}.
Thus, if the cell electrical/thermal properties are known or can be
identified, this information can be exploited to improve
the estimate of the thermal state of the cell or reduce the number of sensors required.

In this study we demonstrate that ITD can be used as the measurement input
to a thermal model in order to estimate the cell temperature distribution.
First, an extended Kalman filter (EKF) is used to estimate the cell temperature distribution when all the relevant thermal parameters, including the convection coefficient, are known. The thermal model consists of a polynomial approximation (PA) to the 1D cylindrical heat equation. The measurement input consists of the cell current and voltage, along with periodic measurements of the real part of
the impedance at a single frequency. Second, a dual extended Kalman
filter (DEKF) is used to identify the convection coefficient online
when the latter is not known. The predicted core and surface temperatures in each case are validated against core and surface thermocouple measurements, with agreement to within 0.47 $^{\circ}$C (2.4\% of the core temperature increase). The performance of the combined thermal model plus ITD estimator is comparable to
the performance of the same thermal model coupled with conventional
surface temperature measurements. Table \ref{tab:comparison of techniques}
shows the present study in the context of existing temperature estimation
techniques, which highlights that this is the first study to use impedance
as the measurement input to a thermal model.
\begin{table}[h]
\begin{centering}
\begin{tabular}{|c|c|c|c|}
\hline 
Study & Model & \multicolumn{2}{c|}{Measurement}\tabularnewline
\hline 
\hline 
 &  & $T_{surf}$ & ITD\tabularnewline
\hline 
Forgez et al. \cite{Forgez2010a} & $\checkmark$ & $\checkmark$ & \tabularnewline
\hline 
Lin et al. \cite{Lin2013f, Lin2014} & $\checkmark$ & $\checkmark$ & \tabularnewline
\hline 
Kim et al. \cite{Kim2013, Kim2014b} & $\checkmark$ & $\checkmark$ & \tabularnewline
\hline 
Srinivasan et al. \cite{Srinivasan2011c, Srinivasan2012a} &  &  & $\checkmark$\tabularnewline
\hline 
Raijmakers et al. \cite{Raijmakers2014d} &  &  & $\checkmark$\tabularnewline
\hline 
Richardson et al. \cite{Richardson2014} &  & $\checkmark$ & $\checkmark$\tabularnewline
\hline 
\textbf{Present study} & $\checkmark$ &  & $\checkmark$\tabularnewline
\hline 
\end{tabular}
\par\end{centering}
\caption{Comparison of online temperature estimation techniques.\label{tab:comparison of techniques}}
\end{table}

\section{Measurement Principle}
\begin{figure}[h]
\begin{centering}
\includegraphics[width=0.6\columnwidth]{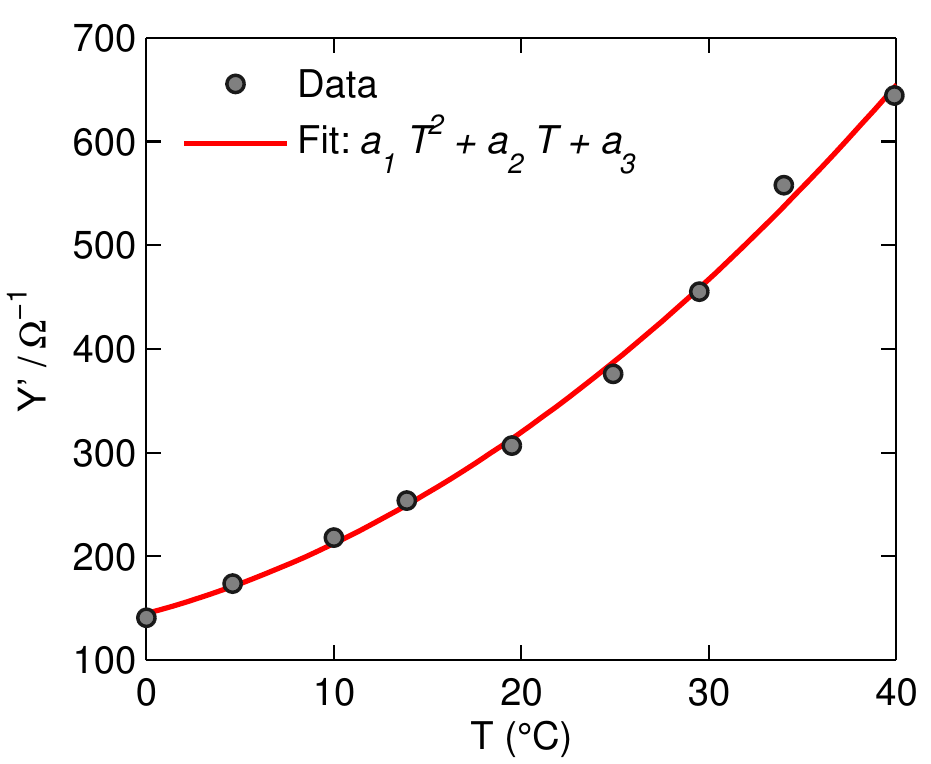}
\par\end{centering}
\caption{Polynomial fit to experimental data of admittance at $f=215\, Hz$
vs. uniform cell temperature.\label{fig:Polynomial-fit}}
\end{figure}

The electrochemical impedance, $Z\left(\omega\right)=Z'\left(\omega\right)+jZ''\left(\omega\right)$, of lithium ion cells is a function of temperature, state of charge
(SOC) and state of health (SOH). Within an appropriate frequency
range, however, the dependence on SOC and SOH is negligible
and the impedance can thus be used to infer information about the
cell temperature \cite{Srinivasan2011c}. Previous ITD studies have
used as a temperature-dependent parameter the real part of the impedance
at a specific frequency \cite{Schmidt2013a}, the phase shift at a
specific frequency \cite{Srinivasan2011c,Srinivasan2012a}, and the
intercept frequency \cite{Raijmakers2014d}. To demonstrate our technique
we use the real part of the impedance at $f$ = 215 Hz. Our previous
work showed that the real part of the cell admittance (the inverse
of the cell impedance) at 215 Hz can be related to the temperature
distribution using a second order polynomial fit. For an annular cell
with inner radius $r_{i}$ and outer radius $r_{o}$ the real part  of the admittance
is given by \cite{Richardson2014}:
\begin{equation}
Y'=\frac{2}{r_{o}^{2}}\int_{r_{i}}^{r_{o}}r\left(a_{1}+a_{2}T(r)+a_{3}T^{2}(r)\right)dr\label{eq:G}
\end{equation}
where $a_{1}$, $a_{2}$ and $a_{3}$ are the 1st, 2nd and 3rd coefficients
of the polynomial relating impedance to uniform cell temperature ($Y'=a_{1}+a_{2}T_{uniform}+a_{3}T_{uniform}^{2}$),
provided that the admittivity varies in the radial direction only.
This assumption is valid if the heat transfer from the top and bottom
ends of the cell is negligible, which is approximately true for cylindrical
cells connected in series with identical cells on either end \cite{Fleckenstein2011},
a configuration which may apply to the majority of cells in a large
battery pack.
However, the application of this approach to cooling configurations involving substantial end cooling would require a more involved expression for the admittance than  eq. \ref{eq:G}, as well as an appropriate modification to the 1D thermal model described in the following section.
%Moreover, although the method is applied to a cylindrical cell, the proposed approach could be generalised to other geometries by using alternative thermal/impedance models.
Moreover, although the method is applied to a cylindrical cell, the proposed approach could be applied to other geometries in a similar fashion.
The polynomial fit (Fig.\ \ref{fig:Polynomial-fit})
was obtained by offline impedance measurements on the cell at multiple
uniform temperatures \cite{Richardson2014}.

\begin{figure}[h]
	{\flushleft{\large\textbf{ a}} %
		\hspace{4cm} {\large\textbf{b}} %
		\par\vspace{0.0cm}}
\begin{centering}
\includegraphics[width=1\columnwidth]{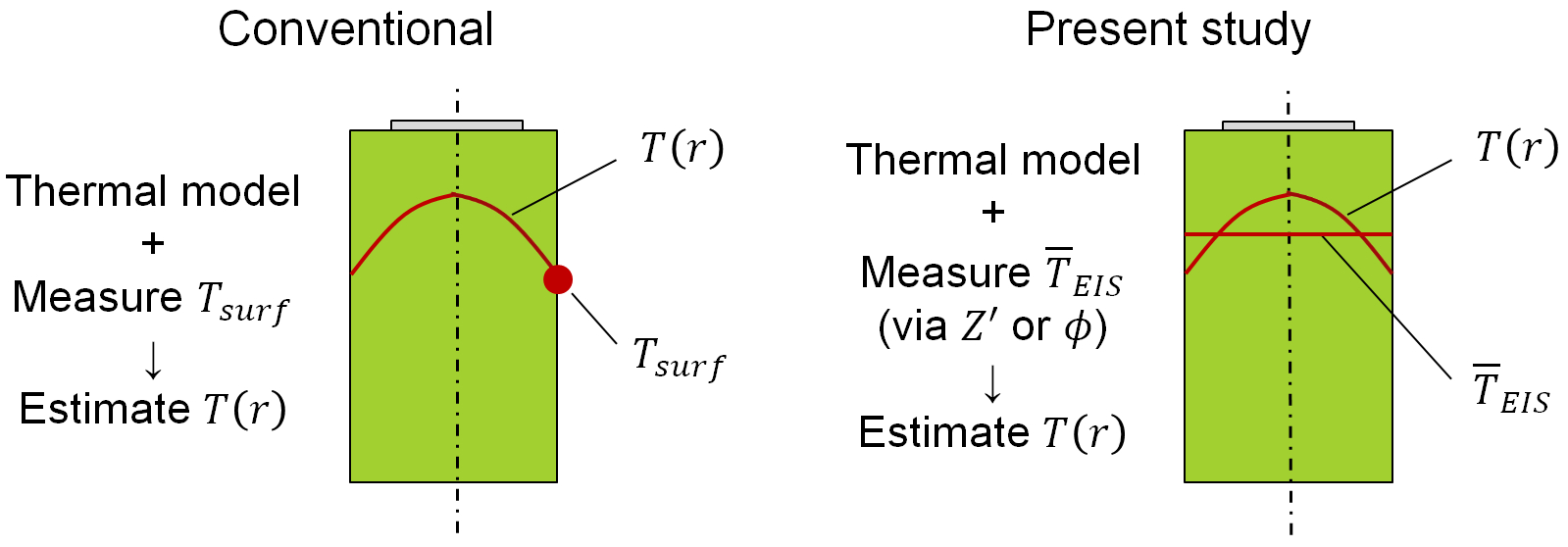}
\par\end{centering}
\caption{Schematic of (a) the conventional approach to temperature estimation and (b)
the proposed approach based on ITD.\label{fig:Schematic}}
\end{figure}

ITD can be viewed as identifying an EIS-based volume average temperature
$\overline{T}_{EIS}$, which is defined as the uniform cell temperature
that would give rise to the measured EIS%
\footnote{Note that, since the impedance temperature relationship is non-linear (as demonstrated in \cite{Troxler2013}), the EIS-based volume average temperature,
$\overline{T}_{EIS}$, is not necessarily equal to the volume average
temperature, $\overline{T}$. Although, it should also be noted that the two are typically close
in value, particularly if the temperature variation within the cell
is small. %
}. Thus the impedance input is similar to a conventional
temperature measurement since it is a scalar function of the internal
temperature distribution but does not uniquely identify the temperature
distribution. Either measurement can therefore be used in conjunction
with a thermal model, as shown in Fig.\ \ref{fig:Schematic}, to estimate core temperature.

\section{Thermal-Impedance Model}

\subsection{Thermal Model}
The cell thermal model consists of the heat equation for 1D unsteady heat conduction in a cylinder, given by the following Boundary Value Problem (BVP) {[}2{]}: \addtocounter{equation}{0}
\begin{subequations}
\begin{equation}
\rho c_{p}\frac{\partial T(r,t)}{\partial t}=k_{t}\frac{\partial^{2}T(r,t)}{\partial r^{2}}+\frac{k_{t}}{r}\frac{\partial T(r,t)}{\partial r}+\frac{Q(t)}{V_{b}}\label{eq:heat_time}
\end{equation}
where $\rho$, $c_p$ and $k_t$ are the density, specific heat capacity and thermal conductivity respectively, $V_b$ is the cell volume, and $Q$ is the heat generation rate. 
The boundary conditions are given by:
\begin{eqnarray}
\left.\frac{\partial T(r,t)}{\partial r}\right|_{r=r_{o}} & = & -\frac{h}{k_{t}}(T(r_{o},t)-T_{\infty}(t)) \label{eq:BC1}\\
\left.\frac{\partial T(r,t)}{\partial r}\right|_{r=0} & = & 0\label{eq:BC2}
\end{eqnarray}
\end{subequations}
where $T_{\infty}$ is the temperature of the heat transfer fluid, and $h$ is the convection coefficient.
A commonly employed expression for the heat source in a lithium ion
battery is
\begin{align}
Q=I(V-U_{OCV})+IT\frac{\partial U_{OCV}}{\partial T}
\label{eq:Q}
\end{align}
which is a simplified version of the expression first proposed by
Bernardi et al \cite{Bernardi1985}. The first term is the heat generation due to ohmic losses in the cell, charge transfer overpotential and
mass transfer limitations. The current $I$ and voltage $V$ for this expression are measured online. The open circuit voltage
$U_{OCV}$ is a function of SOC but is approximated here as a constant value measured at 50 \% SOC, since the HEV drive cycles employed in
this study operate the cell within a small range of SOC ($47 - 63 \, \%$) and therefore OCV variation.
%If necessary, an estimator for $U_{OCV}$ could also be constructed, but we neglect this in the current work.
If necessary, an estimator of $U_{OCV}$ could also be constructed (for example using a dynamic electrical model \cite{Birkl2013a}), but for clarity and brevity we neglect this here.
The second term, the entropic
heat, is neglected in this study because (i) the term $\partial U_{avg} / \partial T$
is small ($0 < \partial U_{avg} / \partial T <0.1$ mVK\textsuperscript{-1}) within the operated range of SOC \cite{Forgez2010a}, and (ii) the net reversible heat would be close to zero when the cell
is operating in HEV mode.
\subsection{Polynomial Approximation}
A polynomial approximation (PA) is used to approximate the solution of eq.\ \ref{eq:heat_time}. The approximation was first introduced in  \cite{Kim2013} and is described in detail in that article, although the essential elements are repeated
here for completeness.

The model assumes a temperature distribution of the form
\begin{align}
T(r,t)=a(t)+b(t)\left(\frac{r}{r_o}\right)^{2}+d(t)\left(\frac{r}{r_o}\right)^{4}
\end{align}
The two states of the model are the volume averaged temperature $\overline{T}$ and
temperature gradient $\overline{\gamma}$:
\begin{align}
\overline{T}=\frac{2}{r_{o}^{2}}\int_{0}^{r_{o}}rTdr,\qquad
\overline{\gamma}=\frac{2}{r_{o}^{2}}\int_{0}^{r_{o}}r\left(\frac{\partial T}{\partial r}\right)dr\label{eq:Tbar}
\end{align}
The temperature distribution is expressed as a function of $\overline{T}$, $\overline{\gamma}$, and the cell surface temperature, $T_{surf}$:
\begin{multline}
T(r,t)=4T_{surf}-3\overline{T}-\frac{15r_{o}}{8}\overline{\gamma}\\
+\left[-18T_{surf}+18\overline{T}+\frac{15r_{o}}{2}\overline{\gamma}\right]\left(\frac{r}{r_{o}}\right)^{2}\\
+\left[15T_{surf}-15\overline{T}-\frac{45r_{o}}{8}\overline{\gamma}\right]\left(\frac{r}{r_{o}}\right)^{4}\label{eq:T_r_t}
\end{multline}

Using \ref{eq:BC1}, the surface temperature can be expressed as
\begin{equation}
T_{surf}=\frac{24k_{t}}{24k_{t}+r_{o}h}\overline{T}+\frac{15k_{t}r_{o}}{48k_{t}+2r_{o}h}\overline{\gamma}+\frac{r_{o}h}{24k_{t}+r_{o}h}T_{\infty}\label{eq:Ts}
\end{equation}

By obtaining the volume-average of eq.\
\ref{eq:heat_time} and of its partial derivative with respect to
$r$, a two-state thermal model consisting of two ODEs is obtained:
\begin{equation}
\begin{aligned}\mathbf{\dot{x}} & =\mathbf{Ax}+\mathbf{Bu}\\
\mathbf{y} & =\mathbf{Cx}+\mathbf{Du}
\end{aligned}
\label{eq:State-space-model}
\end{equation}

where $\mathbf{x}=\left[\overline{T}\;\overline{\gamma}\right]^{T}$,
$\mathbf{u}=\left[Q\; T_{\infty}\right]^{T}$ and $\mathbf{y}=\left[T_{core}\; T_{surf}\right]^{T}$
are state, inputs and outputs respectively. The system matrices $\mathbf{A}$,
$\mathbf{B}$, $\mathbf{C}$, and $\mathbf{D}$ are defined as:
\begin{equation}
\begin{aligned}\mathbf{A} & =\left[\begin{array}{cc}
\frac{-48\alpha h}{r_{o}(24k_{t}+r_{o}h)} & \frac{-15\alpha h}{24k_{t}+r_{o}h}\\
\frac{-320\alpha h}{r_{o}^{2}(24k_{t}+r_{o}h)} & \frac{-120\alpha(4k_{t}+r_{o}h)}{r_{o}^{2}(24k_{t}+r_{o}h)}
\end{array}\right]\\
\mathbf{B} & =\left[\begin{array}{cc}
\frac{\alpha}{k_{t}V_{b}} & \frac{48\alpha h}{r_{o}(24k_{t}+r_{o}h)}\\
0 & \frac{320\alpha h}{r_{o}^{2}(24k_{t}+r_{o}h)}
\end{array}\right]\\
\mathbf{C} & =\left[\begin{array}{cc}
\frac{24k_{t}-3r_{o}h}{24k_{t}+r_{o}h} & -\frac{120r_{o}k_{t}+15r_{o}^{2}h}{8(24k_{t}+r_{o}h)}\\
\frac{24k_{t}}{24k_{t}+r_{o}h} & \frac{15r_{o}k_{t}}{48k_{t}+2r_{o}h}
\end{array}\right]\\
\mathbf{D} & =\left[\begin{array}{cc}
0 & \frac{4r_{o}h}{24k_{t}+r_{o}h}\\
0 & \frac{r_{o}h}{24k_{t}+r_{o}h}
\end{array}\right]
\end{aligned}
\label{eq:State-space-model-matrices}
\end{equation}
where $\alpha = k_t/ \rho c_p$ is the cell thermal diffusivity.

\subsection{Impedance Measurement}
Eq.\ \ref{eq:G} applies to an annulus with inner radius $r_{i}$ and
outer radius $r_{o}$. If the inner radius is sufficiently small,
the cell may be treated as a solid cylinder, and eq.\ \ref{eq:G} becomes

\begin{equation}
Y'=\frac{2}{r_{o}^{2}}\int_{0}^{r_{0}}r\left(a_{1}+a_{2}T(r)+a_{3}T^{2}(r)\right)dr\label{eq:G-1}
\end{equation}
Substituting eq.\ \ref{eq:T_r_t} in eq.\ \ref{eq:G-1}, the real admittance
can be expressed as a function of of $T_{surf}$, $\overline{T}$,
and $\overline{\gamma}$
\begin{equation}
\begin{aligned}Y' & = & a_{1}+a_{2}\overline{T}+3a_{3}\overline{T}^{2}+2a_{3}T_{surf}^{2}-4a_{3}\overline{T}T_{surf}\\
 &  & +\frac{15a_{3}r_{o}^{2}\overline{\gamma}^{2}}{32}+\frac{15a_{3}r_{o}\overline{T}\overline{\gamma}}{8}-\frac{15a_{3}r_{o}T_{surf}\overline{\gamma}}{8}
\end{aligned}
\label{eq:G<->Tbar/gamma}
\end{equation}
Noting from eq.\ \ref{eq:Ts} that $T_{surf}$ is itself a function
of $\overline{T}$, $\overline{\gamma}$ and $T_{\infty}$ and the
cell parameters, admittance is ultimately a function of $\overline{T}$ and $\overline{\gamma}$,
along with the known thermal parameters and environmental temperature.
In other words, for known values of $r_{o}$, $k_{t}$,
$c_{p}$, $\rho$, and $h$, the impedance is a function of the cell
state and $T_{\infty}$, thus: 
\begin{equation}
Z'=f(\overline{T},\overline{\gamma},T_{\infty})\label{eq:f}
\end{equation}

\section{Frequency Domain Analysis\label{sec:Frequency-Domain-Analysis}}
In this section we analyze the error associated with the polynomial approximation, by comparing the frequency response of the PA model to the frequency response of a full analytical solution of eq.\ (\ref{eq:heat_time}). We also examine the approximation employed in \cite{Richardson2014}, which used a combination
of impedance and surface temperature measurements but no thermal model.
To achieve a unique solution in that case, it was necessary to impose the assumption of 
a quadratic solution to the temperature distribution, and so
we refer to this here as the `quadratic assumption' (QA) solution. 

As in \cite{Kim2013}, the frequency response function of the PA thermal
system, $\mathbf{H}(s)$, is calculated by
\begin{equation}
\mathbf{H}(s)=\mathbf{D}+\mathbf{C}(s\mathbf{I}-\mathbf{A})^{-1}\mathbf{B}
\end{equation}

where $s=j\omega$ and $\mathbf{I}$ is the identity matrix. The frequency
response of the analytical model is derived in \cite{Muratori2010a}% this stops a space
%. The frequency response of the QA model was derived by modifying the approach in \cite{Muratori2010a}.
, and that of the QA solution is derived in Appendix \ref{sub:Frequency-domain-analysis}.

\begin{figure}[h]
\begin{centering}
\includegraphics[width=1\columnwidth]{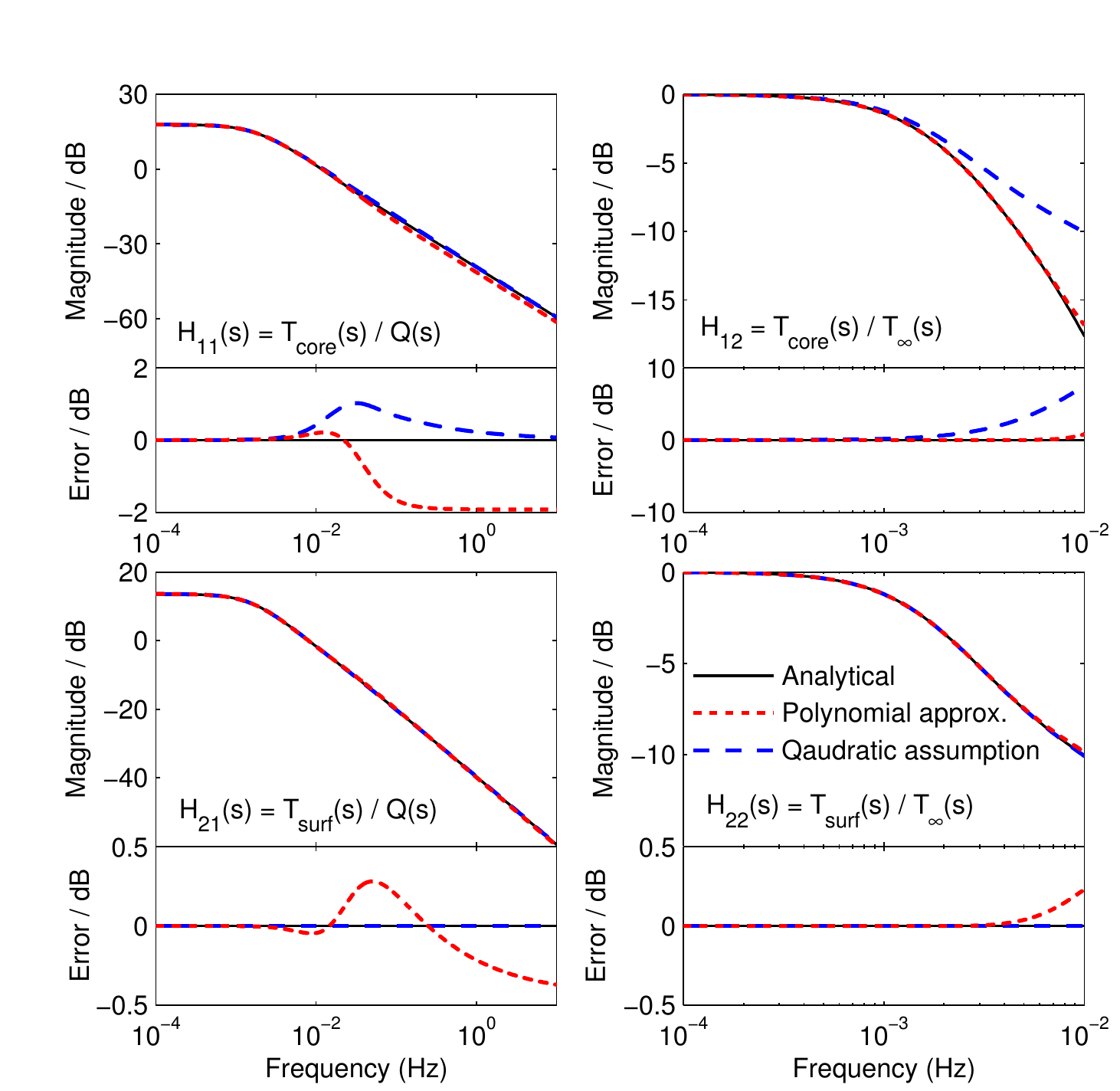}
\par\end{centering}
\caption{Comparison of frequency responses of (i) analytical solution to the
1D cylindrical heat transfer problem, (ii) the polynomial approximation
used in the current study and (iii) the quadratic assumption
used in \cite{Richardson2014}.\label{fig:Frequency-response}}
%\todo[inline]{label corrected}
\end{figure}

Fig.\ \ref{fig:Frequency-response} shows  the impact of changes in heat generation on 
 $T_{core}$ and $T_{surf}$ ($\mathbf{H}_{11}$ and $\mathbf{H}_{21}$
respectively), and the impact of changes in cooling fluid temperature $T_{\infty}$ on $T_{core}$
and $T_{surf}$ ($\mathbf{H}_{12}$ and \textbf{$\mathbf{H}_{22}$}
respectively), for each model along with the error relative to the
analytical solution.
The results in these plots were obtained using the thermal parameters of the 26650 cell used  in the present study (Table \ref{tab:Thermal-parameters}).
The response of $\mathbf{H}_{11}$ and $\mathbf{H}_{21}$
for both the PA and QA models are in good agreement with the analytical
solution (note that the error in the response of $\mathbf{H}_{21}$
and $\mathbf{H}_{22}$ for the QA model is $0$ since it takes the
measured surface temperature as one of its inputs). However, the responses
for both cases to changes in $T_{\infty}$ are less satisfactory.
In particular, the response $\mathbf{H}_{12}$ for the QA model shows
a rapid increase in error relative to the analytical solution at frequencies
above $\sim10^{-3}$ Hz. The PA performs well up to higher frequencies,
although its error in $\mathbf{H}_{21}$ and $\mathbf{H}_{22}$ become
unsatisfactory above $\sim10^{-2}$ Hz. However, the frequency range
at which the PA agrees with the analytical solution is broader than
that of the QA model, and is considered satisfactory given the slow
rate at which the cooling fluid temperature fluctuates in a typical battery thermal
management system.

\section{Experimental}
\begin{figure}[h]
{\Large\textbf{ a}}\par
\begin{centering}
	\includegraphics[width=0.98\columnwidth]{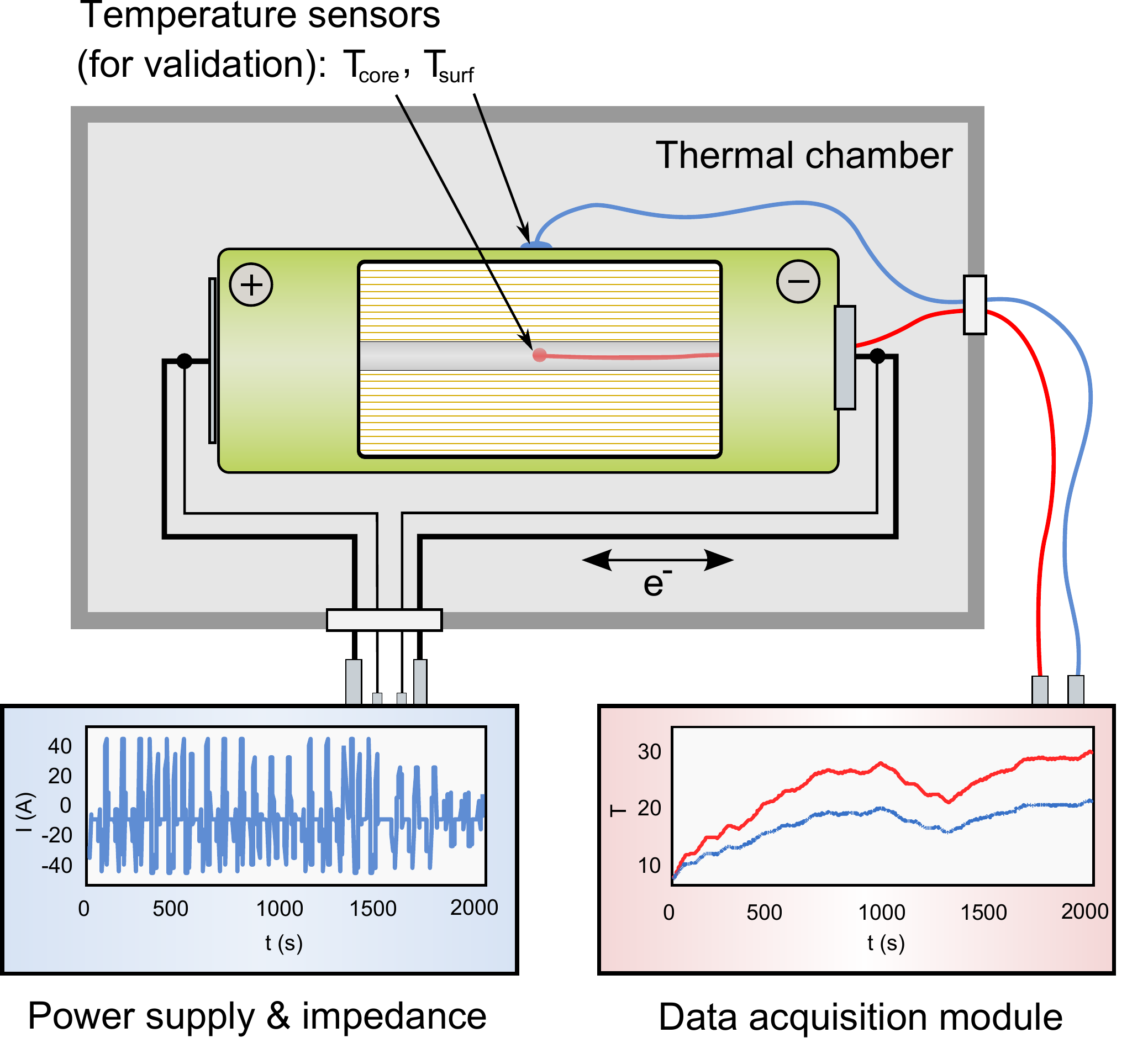}
	\par\end{centering}\vspace{0.0cm}
\begin{centering}
{\flushleft{\Large\textbf{ b}} %
	\hspace{4cm} {\Large\textbf{c}} %
	\par\vspace{0.1cm}}
\includegraphics[height=0.52\columnwidth]{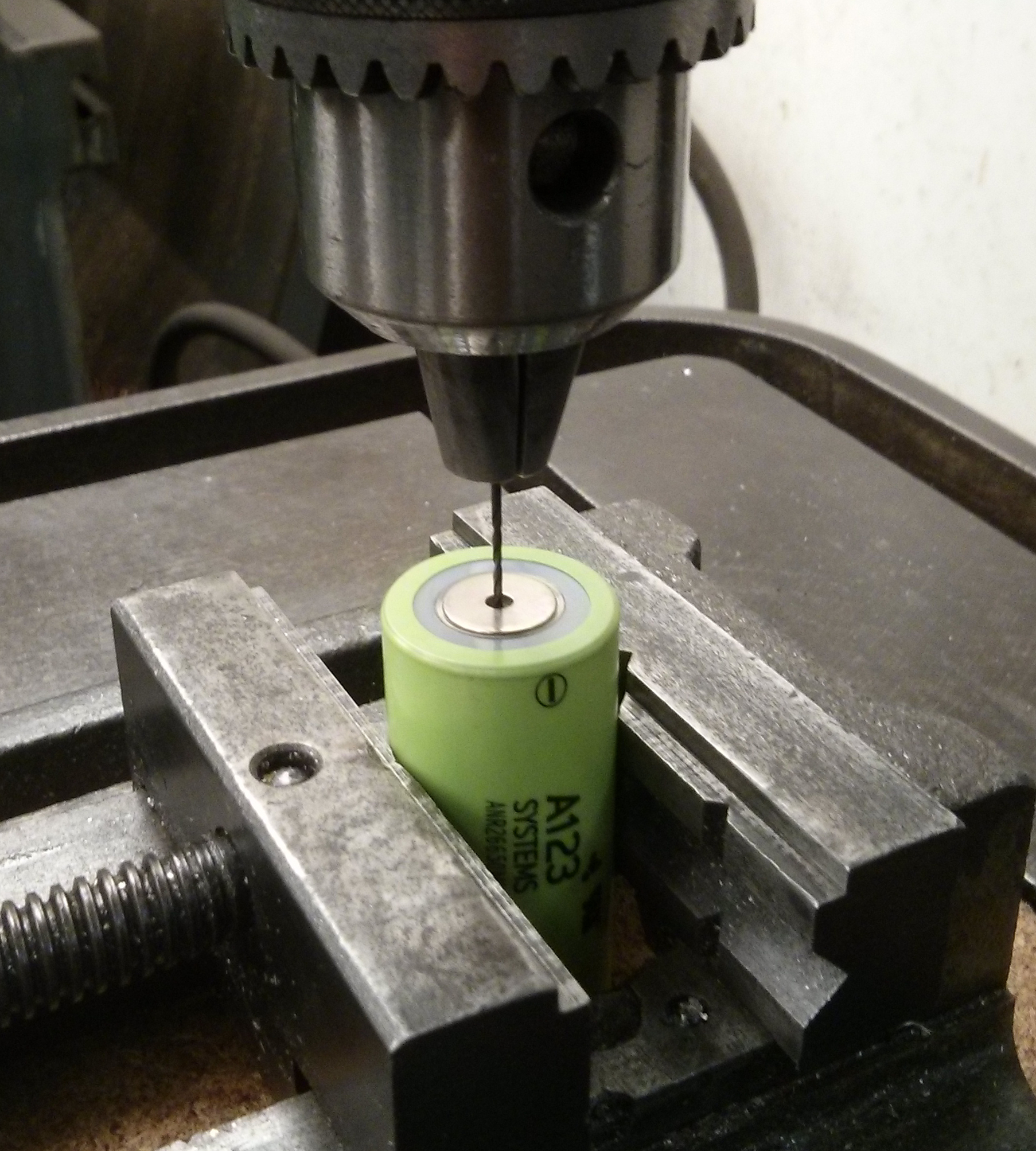} %
\hspace{0.2cm}\includegraphics[height=0.52\columnwidth]{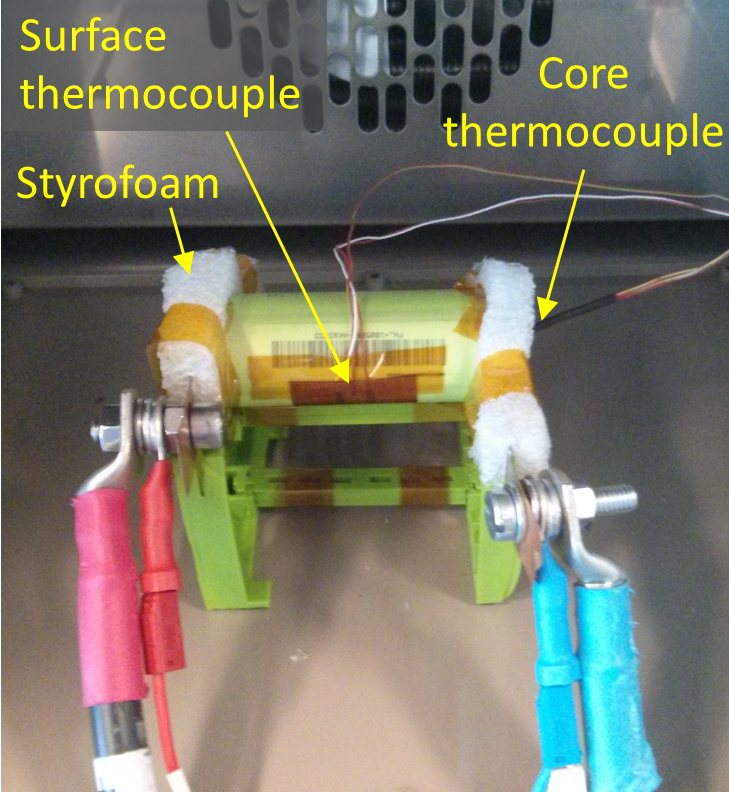} %
{\flushleft{\Large\textbf{ d}} %
	\par\vspace{0.1cm}}
\par\end{centering}
\begin{centering}
\includegraphics[width=0.98\columnwidth]{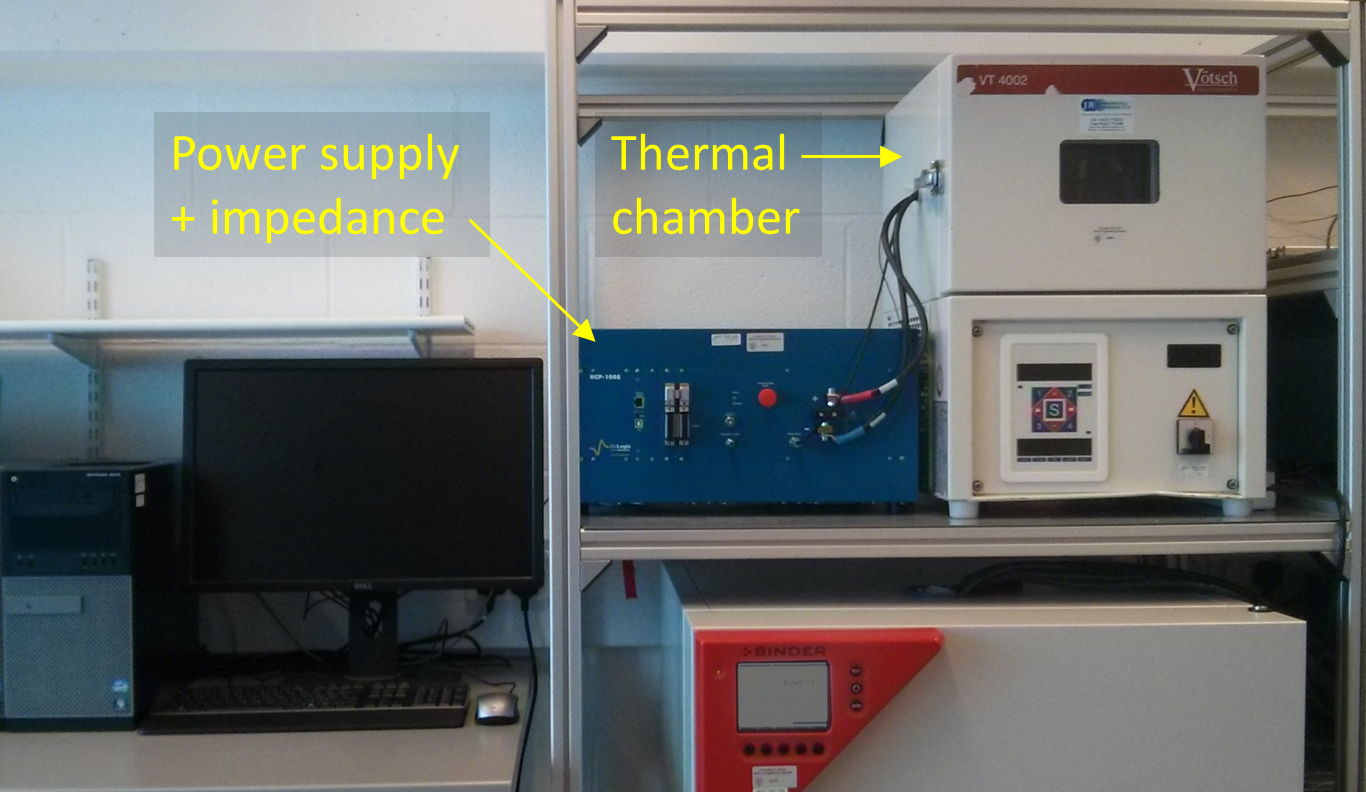}
\par\end{centering}
\caption{Experimental setup for parameter identification and validation. (a) schematic diagram with cutaway view showing cell core and jelly roll, (b) cell drilling procedure, (c) prepared cell inside thermal chamber, (d) power supply \& thermal chamber.} \label{fig:Experimental-setup}
\end{figure}

Experiments were carried out with a 2.3 Ah cylindrical cell (A123
Model ANR26650 m1-A, length 65 mm, diameter 26 mm) with LiFePO$_{4}$ positive electrode and graphite negative electrode.
% The properties of the cell are given in Table \ref{tab:Cell-properties}.
The cell was fitted with two thermocouples, one on the surface and
another inserted into the core via a hole which was drilled in the
positive electrode end (Fig.\ \ref{fig:Experimental-setup}). Cell
cycling and impedance measurements were carried out using a Biologic
HCP-1005 potentiostat/booster. The impedance was measured using Galvanostatic
Impedance Spectroscopy with a 200 mA peak-to-peak perturbation
current. The environmental temperature was controlled with a Votsch
VT4002 thermal chamber. The chamber includes a fan which operates
continuously at a fixed speed during operation.

In order to calibrate the impedance against temperature, EIS measurements
were first conducted on the cell in thermal equilibrium at a
range of temperatures. These experiments and the subsequent identification
of the polynomial coefficients $a_{1}$, $a_{2}$, and $a_{3}$ are
described in \cite{Richardson2014}.

Dynamic experiments were then conducted using two 3500 s current excitation profiles
- the first to parameterise the thermal model, and the second
% (shown in Fig.\ \ref{fig:Current-excitation-profile})
to validate the identified
parameters and to demonstrate the temperature estimation technique.
The profiles were generated by looping over different portions of
an Artemis HEV drive cycle.
The applied currents were in the range $-23$ A to $+30$ A. 
For the duration of the experiments, single
frequency (215 Hz) impedance measurements were carried out every 24
s and the surface and internal temperatures were also monitored. In order
to minimise heat loss through the cell ends, these were insulated using
Styrofoam (Fig.\ \ref{fig:Experimental-setup}). Before each experiment,
the SOC was adjusted to 50\% by drawing a 0.9 C current. The temperature of the thermal chamber was set to 8 $^{\circ}$C and the
cell was allowed to rest until its temperature equilibrated before experiments began.

Besides being a function of temperature and SOC, the impedance is
also a function of DC current, mainly due to the charge transfer polarization
decreasing with increasing current \cite{Ratnakumar2006}. Previous
results confirmed that when the EIS perturbation current is superposed
over an applied DC current, the impedance measurement is altered
\cite{Richardson2014}. To overcome this, the cell was allowed to
rest briefly for 4 s before each EIS measurement was taken, i.e.\ 20 s periods
of the excitation current followed by 4 s rests. The duration of this
rest period was kept as short as possible to ensure that the thermal
response of the cell to the applied cycle was not significantly altered,
and it was found that the core cell temperature dropped by at most
0.25$^{\circ}$C during these rest periods. The issue of impedance measurement
under DC current warrants further investigation.

%\begin{figure}[h]
%	\begin{centering}
%		\includegraphics[width=0.97\columnwidth]{\string"Images/experiment_schematic_box\string".pdf}
%		\par\end{centering}\vspace{0.25cm}
%	%\end{figure}
%	%
%	%\begin{figure}[h]
%	\begin{centering}
%		\includegraphics[height=0.52\columnwidth]{Images/battery_drill_cropped}\hspace{0.2cm}\includegraphics[height=0.52\columnwidth]{Images/cell_annotated}\vspace{0.2cm}
%		\par\end{centering}
%	\begin{centering}
%		\includegraphics[width=0.98\columnwidth]{Images/Equipment_annotated2}
%		\par\end{centering}
%	\caption{Experimental setup for parameter identification and validation experiments.\label{fig:Experimental-setup}}
%\end{figure}

\section{Model Parameterisation \& Validation}
Parameterisation is performed to estimate the values of $k_{t}$,
$c_{p}$, and $h$. The density $\rho$ was identified in advance
by measuring the cell mass and dividing by its volume. The values
from the first excitation profile (comprising cell current, voltage plus surface, core and chamber temperatures)
were used for the estimation.

The parameterisation was carried out offline using \emph{fminsearch} in Matlab to minimise the magnitude of the Euclidean distance between the measured and estimated core and surface temperatures, as in \cite{Kim2014b}.
Table \ref{tab:Thermal-parameters}
compares the identified parameters with the initial guesses and parameters for the same cell from the literature. The estimated parameters are close to those reported
in the literature. The deviations may be attributed to manufacturing
variability, error in the heat generation calculation
(due to the omission of entropic heating in all of these studies),
heat generation in the contact resistances between the cell and connecting
wires and/or measurement uncertainty in the temperature. The convection
coefficient is within the range expected of forced convection air
cooling \cite{Incropera2007a}.

\begin{table}[h]
\begin{centering}
\begin{tabular}{|c|c|>{\centering}p{2cm}|c|c|}
\hline 
Parameter & Units & Reference & Initial & Identified\tabularnewline
\hline 
\hline 
$\rho$ & kg m\textsuperscript{-3} & 2047-2118\linebreak
\cite{Kim2014b,Khasawneh2011a,Li2013a} & - & 2107\tabularnewline
\hline 
$c_{p}$ & J kg\textsuperscript{-1} K\textsuperscript{-1} & 1004.9-1109.2\linebreak\cite{Forgez2010a,Lin2014,Kim2014b} & 1050 & 1171.6\tabularnewline
\hline 
$k_{t}$ & W m\textsuperscript{-1} K\textsuperscript{-1} & 0.488-0.690\linebreak \cite{Muratori2010a,Khasawneh2011a,Kim2014b} & 0.55 & 0.404\tabularnewline
\hline 
$h$ & W m\textsuperscript{-2} & - & 20 & 39.3\tabularnewline
\hline 
\end{tabular}
\par\end{centering}
\caption{Comparison of reference \& estimated parameters\label{tab:Thermal-parameters}}
\end{table}

\begin{figure}[h]
\begin{centering}
\includegraphics[width=1\columnwidth]{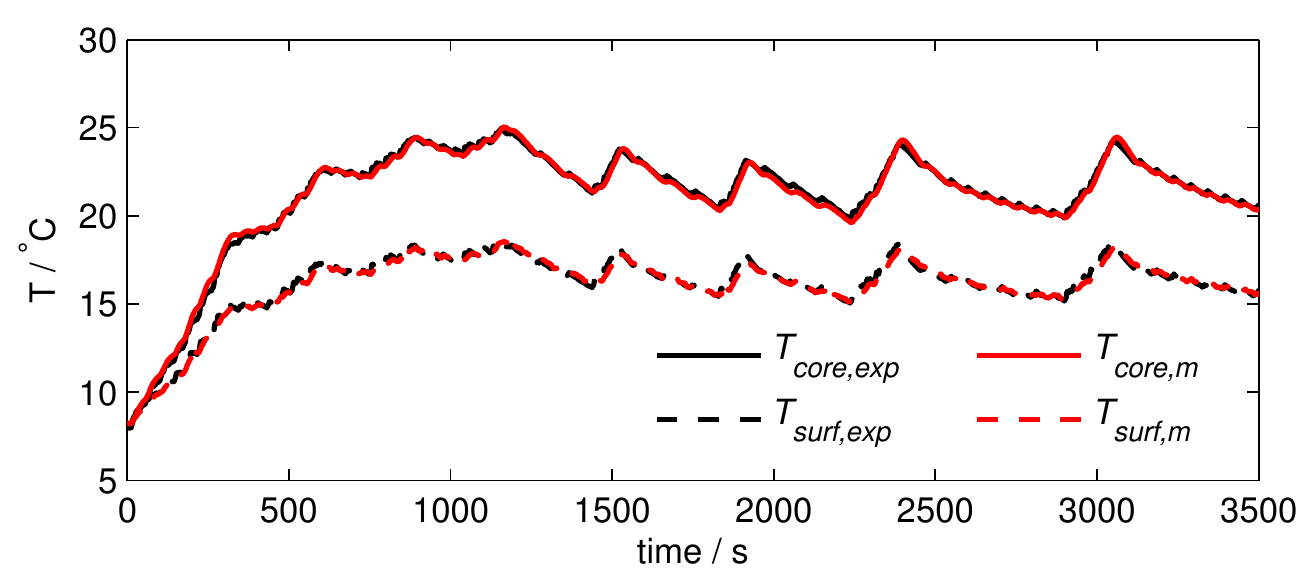}
\par\end{centering}
\caption{Model parameterisation: Comparison between measured and predicted
core and surface temperatures in the parameterised model.\label{fig:Parameterisation}}
\end{figure}

The measured core and surface temperatures (subscript `exp') and the corresponding model predictions (subscript `m') for the parameterised model are shown in Fig.\ \ref{fig:Parameterisation}.
The root-mean-square errors (RMSE) in the surface and core temperatures are 0.19 $^{\circ}$C
and 0.18 $^{\circ}$C respectively. 

\begin{figure}[h]
\begin{centering}
\includegraphics[width=1\columnwidth]{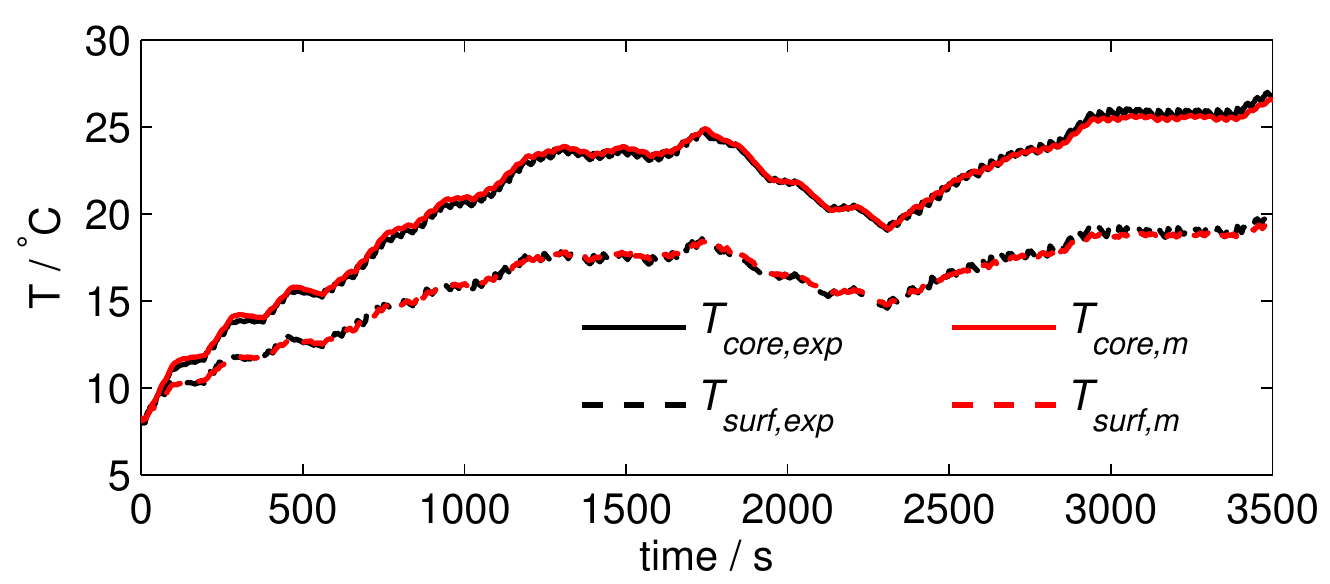}
\par\end{centering}
\caption{Model validation. Comparison between measured and predicted core and
surface temperatures in the parameterised model applied to the second
current excitation profile.\label{fig:Validation}}
\end{figure}

The model with identified parameters was validated against the second
current excitation profile (Fig.\ \ref{fig:Validation}). The RMSEs in the core and surface temperatures were 0.21 $^{\circ}$C
and 0.16 $^{\circ}$C respectively in this case. These errors are only
marginally greater than those in the parameterisation test, indicating
that the estimation is satisfactory.

\section{State Estimation}
Kim et al. \cite{Kim2013} showed that the effect of changes to the
value of $h$ on the predicted surface and core temperatures is greater
than the effect of changes to the other thermal parameters. Moreover, $h$ depends strongly on the thermal
management system settings and its calculation often relies on empirical
correlations between coolant flow rates and heat transfer. Thus, there
is a need to identify the convection coefficient online during operation.
This section outlines the use of a dual extended Kalman filter (DEKF) \cite{Wan2001}
for estimating the core and surface temperatures and the convection
coefficient. The DEKF reduces to an EKF if the convection coefficient
is assumed known and provided to the model in advance.

We firstly modify eq.\ \ref{eq:State-space-model-matrices} by rewriting it as a discrete time model, setting the impedance as the model output, and explicitly including the dependency on the parameter, $h_k$:
\begin{align}
\mathbf{x}_{k+1} & =\bar{\mathbf{A}}(h_{k})\mathbf{x}_{k}+\bar{\mathbf{B}}(h_{k})\mathbf{u}_{k}+\mathbf{v}_{k}\\
y_{k} & =f(\mathbf{x}_{k},\, h_{k})+n_{k}\\
h_{k+1} & =h_{k}+e_{k}\label{eq:param-filt-1}
\end{align}
where $y_{k}=Z'$ and $f(\mathbf{x}_{k},\, h_{k})$ is the non-linear
function relating the state vector to the measurement (i.e. eq.\ \ref{eq:f}), and $\mathbf{v}_{k}$, $n_{k}$ and $e_{k}$ are the noise inputs of the covariance matrices $\mathbf{R}^{\mathbf{v}}$,
$R^{n}$ and $R^{e}$.
The states, inputs and measured outputs are  thus $\mathbf{x}=\left[\overline{T}\;\overline{\gamma}\right]^{T}$,
	$\mathbf{u}=\left[Q\; T_{\infty}\right]^{T}$ and 
	$y=Z'$.
Note that, although the impedance is now the model output, the core and surface temperatures are also computed from the identified states and parameter at each time step using eq.\ \ref{eq:State-space-model}, for validation against the thermocouple measurements.
$\bar{\mathbf{A}}$ and $\bar{\mathbf{B}}$
are system matrices in the discrete-time domain, given by
\begin{equation}
\bar{\mathbf{A}}=e^{(\mathbf{A}\Delta t)},\:\bar{\mathbf{B}}=\mathbf{A}^{-1}(\bar{\mathbf{A}}-\mathbf{I})\mathbf{B}
\end{equation}
where $\Delta t$ is the sampling time of 1 s. The update processes
are then given as follows. The time update processes for the parameter
filter are:
\begin{align}
\hat{h}_{k}^{-} & =\hat{h}_{k-1}\label{eq:param-filt-2}\\
(P_{k}^{h})^{-} & =P_{k-1}^{h}+R^{e}\label{eq:param-filt-3}
\end{align}
where $\hat{h}_{k}^{-}$ and $\hat{h}_{k}$ are the \emph{a priori
}and \emph{a posteriori} estimates of the parameter $h$, and $(P_{k}^{h})^{-}$
and $P_{k-1}^{h}$ are the corresponding\emph{ }error covariances. 

The time update processes for the state filter are:
\begin{align}
\hat{\mathbf{x}}_{k}^{-} & =\bar{\mathbf{A}}_{k-1}\hat{\mathbf{x}}_{k-1}+\bar{\mathbf{B}}_{k-1}\mathbf{u}_{k-1}\\
(\mathbf{P}_{k}^{\mathbf{x}})^{-} & =\bar{\mathbf{A}}_{k-1}\mathbf{P}_{k-1}^{\mathbf{x}}\bar{\mathbf{A}}_{k-1}^{T}+\mathbf{R}^{\mathbf{v}}
\end{align}
where $\mathbf{\hat{x}}_{k}^{-}$ and $\mathbf{\hat{x}}_{k}$ are
the \emph{a priori }and \emph{a posteriori} estimates of the state,
and $(\mathbf{P}_{k}^{\mathbf{x}})^{-}$ and $\mathbf{P}_{k-1}^{\mathbf{x}}$
are the corresponding\emph{ }error covariances. The matrices $\bar{\mathbf{A}}_{k-1}$
and $\bar{\mathbf{B}}_{k-1}$ are calculated by:
\begin{align}
\bar{\mathbf{A}}_{k-1}=\bar{\mathbf{A}}(h)\big|_{h=\hat{h}_{k}^{-}},\,\bar{\mathbf{B}}_{k-1}=\bar{\mathbf{B}}(h)\big|_{h=\hat{h}_{k}^{-}}
\end{align}
Since the relationship between impedance and the cell state is non-linear,
the measurement model must be linearised about the predicted observation
at each measurement. The measurement update equations for the state
filter are:
\begin{align}
\mathbf{K}_{k}^{\mathbf{x}} & =(\mathbf{P}_{k}^{\mathbf{x}})^{-}(\mathbf{H}_{k}^{\mathbf{x}})^{T}\left(\mathbf{H}_{k}^{\mathbf{x}}(\mathbf{P}_{k}^{\mathbf{x}})^{-}\mathbf{(H}_{k}^{\mathbf{x}})^{T}+R^{n}\right)^{-1}\\
\hat{\mathbf{x}}_{k} & =\hat{\mathbf{x}}_{k}^{-}+\mathbf{K}_{k}^{\mathbf{x}}\left(z_{k}-f(\mathbf{\hat{x}}_{k}^{-},\,\hat{h}_{k}^{-})\right)\\
\mathbf{P}_{k}^{\mathbf{x}} & =(\mathbf{I}-\mathbf{K}_{k}^{\mathbf{x}}\mathbf{H}_{k}^{\mathbf{x}})(\mathbf{P}_{k}^{\mathbf{x}})^{-}
\end{align}
where $\mathbf{K}_{k}^{\mathbf{x}}$ is the Kalman gain
for the state,\emph{ }and $\mathbf{H}_{k}^{\mathbf{x}}$ is the Jacobian
matrix of partial derivatives of $f$ with respect to $\mathbf{x}$:
\begin{align}
\mathbf{H}_{k}^{\mathbf{x}}=\frac{\partial f(\mathbf{x}_k,h_k)}{\partial\mathbf{x}_k}\Bigg|_{\mathbf{x}_k=\hat{\mathbf{x}}_{k}^{-}}
\end{align}
The measurement update processes for the parameter filter are:
\begin{align}
K_{k}^{h} & =(P_{k}^{h})^{-}(H_{k}^{h})^{T}\left(H_{k}^{h}(P_{k}^{h})^{-}(H_{k}^{h})^{T}+R^{n}\right)^{-1}\label{eq:param-filt-4}\\
\hat{h}_{k} & =\hat{h}_{k}^{-}+K_{k}^{h}\left(z_{k}-f(\mathbf{\hat{x}}_{k},\,\hat{h}_{k}^{-})\right)\\
P_{k}^{h} & =(I-K_{k}^{h}H_{k}^{h})(P_{k}^{h})^{-}
\end{align}
where $H_{k}^{h}$ is the Jacobian matrix of partial derivatives of
$f$ with respect to $h$:
\begin{align}
H_{k}^{h}=\frac{\partial f(\mathbf{x}_k,h_k)}{\partial h_k}\Bigg|_{ h_k=\hat{h}_{k}^{-}}\label{eq:param-filt-5}
\end{align}

The above algorithm can be simplified to a standard EKF by omitting the parameter update processes (eqs.\ \ref{eq:param-filt-1}, \ref{eq:param-filt-2}-\ref{eq:param-filt-3} and \ref{eq:param-filt-4}-\ref{eq:param-filt-5}) and assuming the convection coefficient is fixed. In the following section we investigate the performance of both the baseline EKF and the full DEKF algorithm.

\section{Results}
We first investigate the performance of an EKF estimator
whereby the convection coefficient is provided and assumed fixed.
We then compare the performance of the DEKF algorithm with that of
the baseline EKF when an incorrect initial estimate of the convection
coefficient is provided. Lastly, we compare the performance of the
DEKF with that of a dual Kalman filter (DKF) based on the same thermal model but with
 $T_{surf}$ as the measurement input rather than $Z'$.

\subsection{Convection Coefficient Known\label{sub:Convection-Coefficient-Provided}}
\begin{figure}%[h]
\begin{raggedright}
\includegraphics[width=0.95\columnwidth]{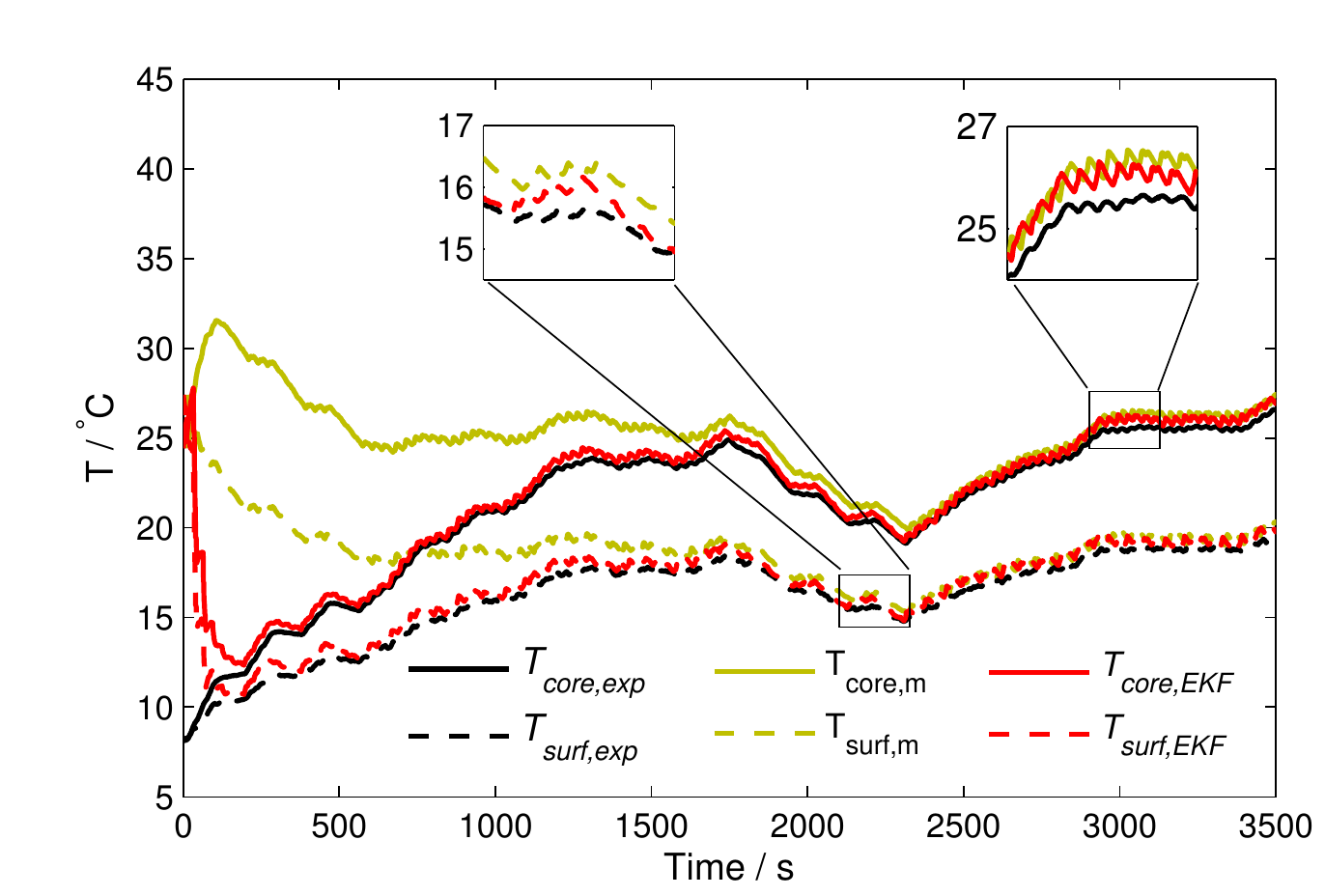}
\par\end{raggedright}
\caption{Temperature results for EKF using $Z'$ as measurement input\label{fig:EKF}}
\end{figure}

The initial state estimate provided to the battery is $\mathbf{\hat{x}_0}=[25\:0]$,
i.e.\ the battery has a uniform temperature distribution at 25 $^{\circ}$C.
The true initial battery state is a uniform temperature distribution
at 8 $^{\circ}$C. The covariance matrices are calculated as $R^{n}=\sigma_{n}^{2}$
and $\mathbf{R}^{\mathbf{v}}=\beta_{v}^{2}diag(2,2)$.
The first tuning parameter is chosen as $\sigma_{n}=1\times10^{-4}$ $\Omega$, which is a rough estimate of the standard deviation of the impedance measurement. The second tuning parameter was chosen as $\beta_{v}=0.1$, by trial and error.
%where the tuning
%parameters are chosen as $\sigma_{n}=1\times10^{-4}\,\Omega$ and
%$\beta_{v}=0.1$. 
Fig.\ \ref{fig:EKF}
shows that, using the EKF, the core and surface temperatures quickly
converge to the correct values and are accurately estimated throughout
the entire excitation profile. The RMSEs of core and surface temperature
are 1.35 $^{\circ}$C and 1.34 $^{\circ}$C respectively. In contrast,
the RMSEs for the open loop model (subscript `m') with no measurement feedback are
6.66 $^{\circ}$C and 4.42 $^{\circ}$C respectively.
It should be noted that since the uncertainty of the impedance measurement typically increases as impedance decreases, the temperature estimates could be more uncertain at higher temperatures. Hence, the implementation of this technique could be more challenging at higher ambient temperature conditions than those studied here.

It should be noted that we also achieved similar performance using
a simpler EKF based on $Z'$ with the assumption that the impedance
is related directly to $\overline{T}$ rather than to 
$\overline{T}_{EIS}$. However, this assumption may lead to unsatisfactory
results for cells with a larger radius or when larger temperature
gradients exist within the cell. Moreover, since this approach assumes that the impedance is a function of the state only (and not the parameter $h$), it is not suitable for the application of the DEKF discussed in the following section.

\subsection{Convection Coefficient Unknown\label{sub:Convection-Coefficient-Unkown}}
\begin{figure}%[h]
\begin{centering}
\includegraphics[width=1\columnwidth]{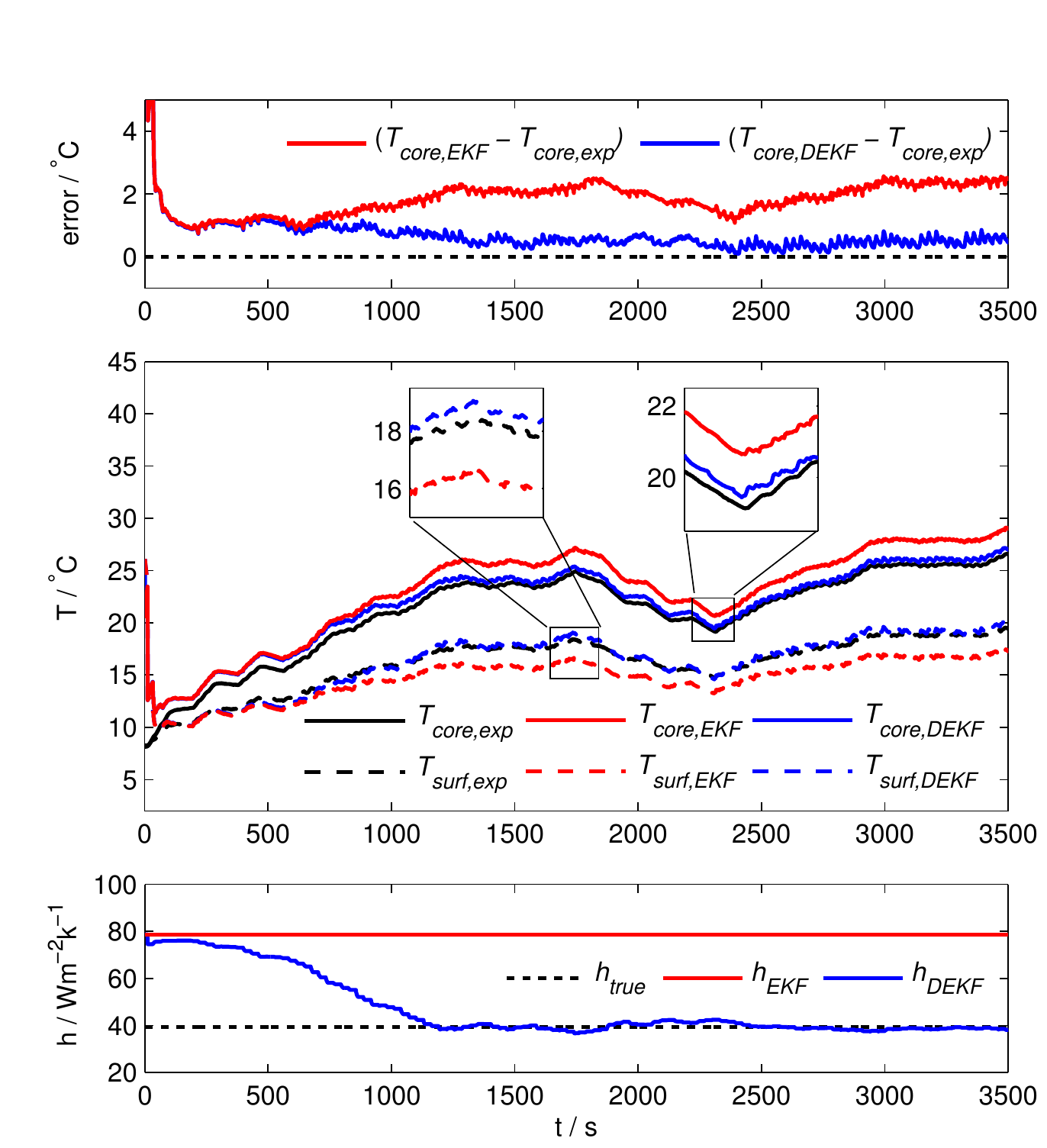}
\par\end{centering}
\caption{Temperature results for DEKF using $Z'$ as measurement input\label{fig:DEKF-with-Z}}
\end{figure}

Next we investigate the performance of the DEKF. The same incorrect
initial state estimate is provided to the battery, $\hat{x}_0=[25\:0]$.
Moreover, an incorrect initial estimate for the convection coefficient
is provided, $\hat{h}_0=2\times h_{true}$. This value of $h$ is also
provided to the EKF. The error covariance matrix for the parameter
estimator is $R^{e}=\beta_{e}^{2}$ where the tuning parameter is
chosen as $\beta_{e}=2.5$. Fig.\ \ref{fig:DEKF-with-Z} compares the
results of both of these cases against the thermocouple measurements.
The EKF is shown to overestimate the core temperature and underestimate
the surface temperature for the duration of the experiments. This
is expected, since the impedance measurement ensures the
accuracy of the volume averaged cell temperature but the model assumes that the convection coefficient is higher than in reality, and therefore 
the temperature difference across the cell is overestimated. In contrast,
the DEKF corrects the convection coefficient, and thus improves the
accuracy of the subsequent temperature predictions. This is evident
from the errors in the core temperature estimate (top plot of Fig.\
\ref{fig:DEKF-with-Z}), which initially are similar in both cases
but drop to much smaller values for the DEKF once the correct convection
coefficient is identified. The RMSEs of core and surface temperature
in each case are shown in Table \ref{tab:Comparison-of-RMSEs}, along
with the values for the time period, 1200 < t < 3500 s (i.e. after
the convection coefficient value has converged). 

\begin{figure}[h]
\begin{centering}
\includegraphics[width=1\columnwidth]{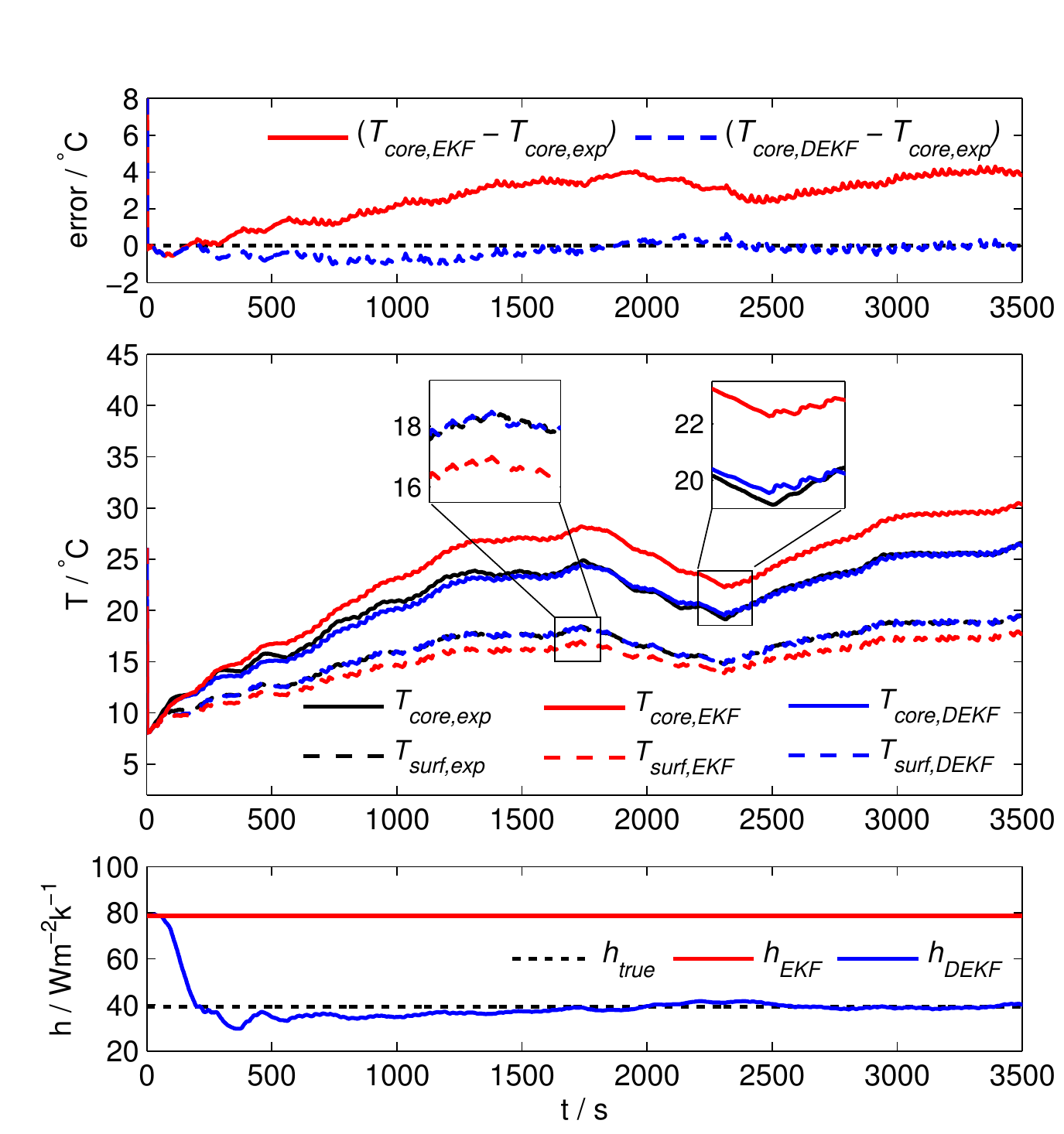}
\par\end{centering}
\caption{DKF using $T_{surf}$ as measurement input.\label{fig:DEKF-using-T_surf}}
\end{figure}

Finally, we investigate the case where $T_{surf}$ is used as the
measurement ($y = T_{surf}$) to the estimator rather than $Z'$. This results
in a linear KF and DKF exactly equivalent to that studied in \cite{Kim2014b}.
The tuning parameters for the covariance matrices are also chosen
to be the same as those employed in \cite{Kim2014b}. The same initial
state and parameter estimates are provided as for the DEKF. Fig.\ \ref{fig:DEKF-using-T_surf}
shows that the standard EKF in this case overpredicts the core temperature
by a greater margin than the EKF based on $Z'$, although the surface
temperature estimate is much more accurate. This is because the surface
thermocouple ensures an accurate surface temperature estimate
and to reconcile this with the overestimated convection coefficient,
the core temperature estimate is forced to be much greater. 
The DKF correctly identifies the correct convection coefficient, in the same way as the DEKF. Since
the thermocouple measurement exhibits less noise than the impedance
measurement, the model converges to the correct estimate for $h$
more quickly than in the DEFK case, as shown by the RMSE values
in Table \ref{tab:Comparison-of-RMSEs}.

\begin{table}[h]
\begin{centering}
\begin{tabular}{|c|c|c|c|c|}
\hline 
Method & \multicolumn{2}{c|}{$0<t<3500\,s$} & \multicolumn{2}{c|}{$1200<t<3500\,s$}\tabularnewline
\hline 
 & $T_{core}$ & $T_{surf}$ & $T_{core}$ & $T_{surf}$\tabularnewline
\hline 
\hline 
EKF + $Z'$ & 2.04 & 2.06 & 1.79 & 1.98\tabularnewline
\hline 
KF + $T_{surf}$ & 2.49 & 1.44 & 2.90 & 1.58\tabularnewline
\hline 
DEKF + $Z'$ & 1.43 & 1.24 & 0.47 & 0.42\tabularnewline
\hline 
DKF + $T_{surf}$ & 0.36 & 0.33 & 0.16 & 0.14\tabularnewline
\hline 
\end{tabular}
\par\end{centering}
\begin{centering}
\caption{Comparison of RMSEs for core and surface temperatures (\degree C) with unknown
convection coefficient.\label{tab:Comparison-of-RMSEs}}
\par\end{centering}
\end{table}
In conclusion, the temperature and convection coefficient estimators
using $Z'$ as measurement input are capable of accurately estimating
the core and surface temperatures and the convection coefficient.
The performance is comparable to that of an estimator using the same
thermal model coupled with surface temperature measurements.
Moreover, the performance of the present method is superior to that of previous methods based on impedance measurements alone, which only provide an estimate of the average internal temperature of the cell.

\section{Conclusions}
Impedance temperature detection enables both core and surface temperature estimation without using temperature sensors. In this study, the use of ITD as measurement input to a cell
thermal model is demonstrated for the first time.

Previously, we estimated cell temperature distribution by combining
ITD with a surface temperature measurement and imposing a quadratic
assumption on the radial temperature profile. Frequency domain analysis shows that the QA solution may
be inaccurate if the temperature of the cooling fluid has fluctuations on the order of $10^{-3}$ Hz or higher. The PA thermal
model used in the present study is robust to higher frequency fluctuations ($\sim10^{-2}$ Hz).

An EKF using a parameterised PA thermal model with ITD measurement
input is shown to accurately predict core and surface temperatures
for a current excitation profile based on an Artemis HEV drive cycle.
A DEKF based on the same thermal model and measurement input is capable
of accurately identifying the convection coefficient when the latter
is not provided to the model in advance. The performance of the DEKF
using impedance as measurement input is comparable to an equivalent
DKF estimator using surface temperature as measurement input, although
the latter is slightly superior due to the higher accuracy of the thermocouple.

Future work will investigate the application of ITD to multiple cells
in a battery pack, as well as investigating methods of combining impedance
with conventional sensors to enable more robust temperature monitoring
and fault detection, and self-calibration.

\section*{Acknowledgements}
This work was funded by a NUI Travelling
Scholarship, a UK EPSRC Doctoral Training Award and the Foley-Bejar scholarship from Balliol
College, University of Oxford. This publication also benefited from
equipment funded by the John Fell Oxford University Press (OUP) Research
Fund.
Finally, the authors would like to thank Peter Ireland and Adrien Bizeray for valuable comments, and Robin Vincent for the facilities to prepare the instrumented battery.

\appendix{} %
\subsection{Frequency Domain Analysis of Quadratic Assumption\label{sub:Frequency-domain-analysis}}
The analysis leading to the frequency response plots
of the QA model in Fig.\ \ref{fig:Frequency-response} is outlined in this section. 
The QA model was used in \cite{Richardson2014} to obtain a unique solution for the temperature distribution when the impedance and surface temperature were measured but the cell thermal properties and heat generation rates were assumed unknown. To achieve a unique solution in that case, it was necessary to impose the assumption of a quadratic temperature distribution based on the solution of the 1D heat equation at steady state.

Muratori et al. \cite{Muratori2010a} showed that the time domain PDE of
eq.\ \ref{eq:heat_time}, can be transformed into an equivalent ODE
problem in the frequency domain which can be solved analytically.
Using this approach the solution for the temperatures at the core and surface of the cell are shown to be:
\begin{align}
T_{core}(s) & =\frac{1}{k_t a^{2}}Q(s)+\frac{\frac{h}{k_{t}}(T_{\infty}(s)-\frac{1}{ka^{2}}Q(s))}{\frac{h}{k_{t}}J_{0}(ar_{o})-aJ_{1}(ar_{o})}\label{eq:heat_complex_r_0}\\
T_{surf}(s) & =\frac{1}{k_t a^{2}}Q(s)+\frac{\frac{h}{k_{t}}(T_{\infty}(s)-\frac{1}{k_{t}a^{2}}Q(s))}{\frac{h}{k_{t}}J_{0}(ar_{o})-aJ_{1}(ar_{o})}J_{0}(ar_{o})\label{eq:heat_complex_r_R}
\end{align}
where $a^{2}=s\alpha^{-1}$, $Q(s)$ is the transform of $Q(t)$,
and $J_{i}$ is the $i^{th}$-order Bessel function of the first kind \cite{Kreyszig2010}. Eqs. \ref{eq:heat_complex_r_0} and \ref{eq:heat_complex_r_R} can be interpreted as the outputs of a continuous time dynamic system \cite{Muratori2010a}, where $\mathbf{u}(t)=[Q(t),\, T_{\infty}(t)]^{T}$ and $\mathbf{y}(t)=[T_{core}(t),\, T_{surf}(t)]^{T}$,
such that the solution of the BVP is equivalent to the impulse response
of the system:
\begin{equation}
\begin{bmatrix}T_{core}(s)\\
T_{surf}(s)
\end{bmatrix}=\begin{bmatrix}\mathbf{H}_{11}(s) & \mathbf{H}_{12}(s)\\
\mathbf{H}_{21}(s) & \mathbf{H}_{22}(s)
\end{bmatrix}\begin{bmatrix}Q(s)\\
T_{\infty}(s)
\end{bmatrix}
\end{equation}
where the $\mathbf{H}$ matrix is formed by the transfer functions:
\begin{align}
\mathbf{H}_{11}(s) & =\frac{1}{k_{t} a^2}\frac{\frac{h}{k_{t}}J_{0}(a r_o)-aJ_{1}(ar_o)-\frac{h}{k_{t}}}{\frac{h}{k_{t}}J_{0}(ar_o)-aJ_{1}(ar_o)}\\
\mathbf{H}_{12}(s) & =\frac{\frac{h}{k_{t}}}{\frac{h}{k_{t}}J_{0}(ar_o)-aJ_{1}(ar_o)}\\
\mathbf{H}_{21}(s) & =\frac{1}{k_{t}a^2}\frac{-aJ_{1}(ar_o)}{\frac{h}{k_{t}}J_{0}(ar_o)-aJ_{1}(ar_o)}\\
\mathbf{H}_{22}(s) & =\frac{\frac{h}{k_{t}}J_{0}(ar_o)}{\frac{h}{k_{t}}J_{0}(ar_o)-aJ_{1}(ar_o)}
\end{align}
This system of transfer
functions gives frequency responses for the analytical solution results
plotted in Fig.\ \ref{fig:Frequency-response}. Using a similar approach, we
can develop an analytical solution to the
QA model used in \cite{Richardson2014}, where we denote the new transfer
function $\mathbf{H}_{QA}(s)$. In this case, both the volume-averaged cell
temperature (identified via the impedance%
\footnote{As discussed in Section \ref{sec:Introduction}, the impedance actually
identifies $\overline{T}_{EIS}$, which is not necessarily equal to
$\overline{T}$. However, the assumption that $\overline{T}_{EIS}=\overline{T}$
is satisfactory for the purpose of identifying the frequencies at
which errors relative to the analytical solution become significant, and is only used for this purpose in this article.}) and the surface temperature
were measured directly, and it is assumed that no other information
was available. Since these two inputs alone are not sufficient to
achieve a unique solution for the temperature distribution, it was
also necessary to impose the following constraint on the temperature
profile:
\begin{equation}
T_{QA}(r)=T_{surf}+(T_{core,\, QA}-T_{surf})\left(1-\frac{r^{2}}{r_{o}^{2}}\right)\label{eq:T_QA_r}
\end{equation}
This is the 1D steady-state solution of the heat equation
for a cylinder with uniform heat generation \cite{Incropera2007a},
with $T(r)$ and $T_{core}$ replaced by $T_{QA}(r)$ and $T_{core,\, QA}$. 

The volume average of the QA temperature distribution is set equal to the true volume average temperature, giving:
\begin{equation}
\overline{T}_{QA}=\frac{2}{r_{o}^{2}}\int_{0}^{r_{o}}rT_{QA}(r)\mathrm{d}r=\overline{T}\label{eq:T_QA_bar}
\end{equation}
Substituting eq. \ref{eq:T_QA_r} in eq. \ref{eq:T_QA_bar} and integrating, the QA core temperature
becomes
\begin{equation}
T_{core,\, QA}=2\overline{T}-T_{surf}\label{eq:PSS_T_c}
\end{equation}
Thus, an estimate for the core temperature is obtained directly from
the surface and volume averaged temperature measurements.
Since the surface temperature is measured, the values of $\mathbf{H}_{QA,\,21}(s)$
and $\mathbf{H}_{QA,\,22}(s)$ are identical to the corresponding
values of the unsteady thermal model. The values of $\mathbf{H}_{QA,\,11}(s)$
and $\mathbf{H}_{QA,\,12}(s)$ can be obtained as follows: 
Substituting $\overline{T}$ from eq.\ \ref{eq:Tbar} into eq.\ \ref{eq:PSS_T_c}, the QA approximation of the core temperature can be expressed as a function of the temperature distribution:
\begin{equation}
T_{core,QA}=\frac{4}{r_{o}^{2}}\int_{0}^{r_{o}}\left(rT(r)\mathrm{d}r-T_{surf} \right) dr
\end{equation}
Substituting eqs. \ref{eq:heat_complex_r_0} and \ref{eq:heat_complex_r_R},
for $T(r)$ and $T_{surf}$, respectively, and integrating (noting
that $\int_{0}^{r_{o}}rJ_{0}(ar)\mathrm{d}r=r_{o}J_{1}(ar_{o})/a$),
we obtain:
\begin{multline}
T_{core,\, QA}=\frac{Q(s)}{k_{t}a^{2}}+\\
\frac{\frac{h}{k_{t}}(T_{\infty}(s)-\frac{1}{k_{t}a^{2}}Q(s))}{\frac{h}{k_{t}}J_{0}(ar_{o})-aJ_{1}(ar_{o})}\left[\frac{4J_{1}(ar_{o})}{ar_{o}}-J_{0}(ar_{o})\right]
\end{multline}
Thus, the QA system model is given by:
\begin{equation}
\begin{bmatrix}T_{core,\, QA}(s)\\
T_{surf,\, QA}(s)
\end{bmatrix}=\begin{bmatrix}\mathbf{H}_{QA,\,11}(s) & \mathbf{H}_{QA,\,12}(s)\\
\mathbf{H}_{QA,\,21}(s) & \mathbf{H}_{QA,\,22}(s)
\end{bmatrix}\begin{bmatrix}Q(s)\\
T_{\infty}(s)
\end{bmatrix}
\end{equation}
where the $\mathbf{H}$ matrix is formed by the functions:
\begin{align}
\mathbf{H}_{QA,\,11}(s) & =-\frac{k_{t} a^2 r_{o} J_{1}(a r_o)-2hr_{o} a J_{0}(a r_o)+4hJ_{1}(a r_o)}{k_{t}r_{o} a^3 \left(hJ_{0}(a r_o)-k_{t} a J_{1}(a r_o)\right)}\\
\mathbf{H}_{QA,\,21}(s) & =\frac{-hk_{t} a^3 J_{0}a r_o+4hk_{t} a^2 J_{1}a r_o}{k_{t}r_{o} a^3 \left(hJ_{0}a r_o-k_{t} a J_{1}(a r_o)\right)}\\
\mathbf{H}_{QA,\,12}(s) & =\mathbf{H}_{21}(s)\\
\mathbf{H}_{QA,\,22}(s) & =\mathbf{H}_{22}(s)
\end{align}

\bibliographystyle{IEEEtran}
\bibliography{IEEEabrv,Paper_References}

% \begin{IEEEbiography}[{\includegraphics[width=1in,height=1.25in,clip,keepaspectratio]{./robert_richardson}}]
\begin{IEEEbiography}[{\includegraphics[width=1in,height=1.25in,clip,keepaspectratio]{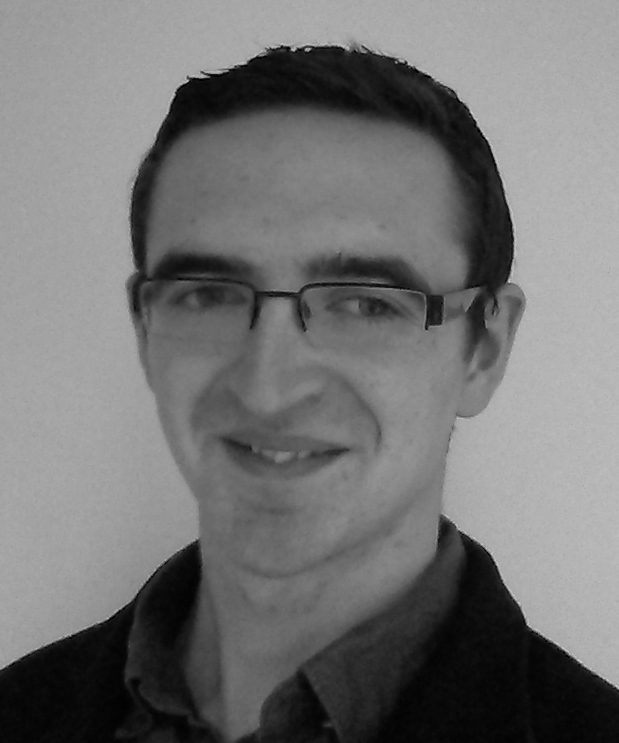}}]
{Robert R. Richardson}
received the B.Eng. degree in mechanical engineering from the National University of Ireland, Galway, in 2012. He is currently working toward the D.Phil (PhD) degree in the Energy and Power Group, Department of Engineering Science, University of Oxford, Oxford, U.K, where his research is focused on impedance based battery thermal management.
\end{IEEEbiography}

\begin{IEEEbiography}[{\includegraphics[width=1in,height=1.25in,clip,keepaspectratio]{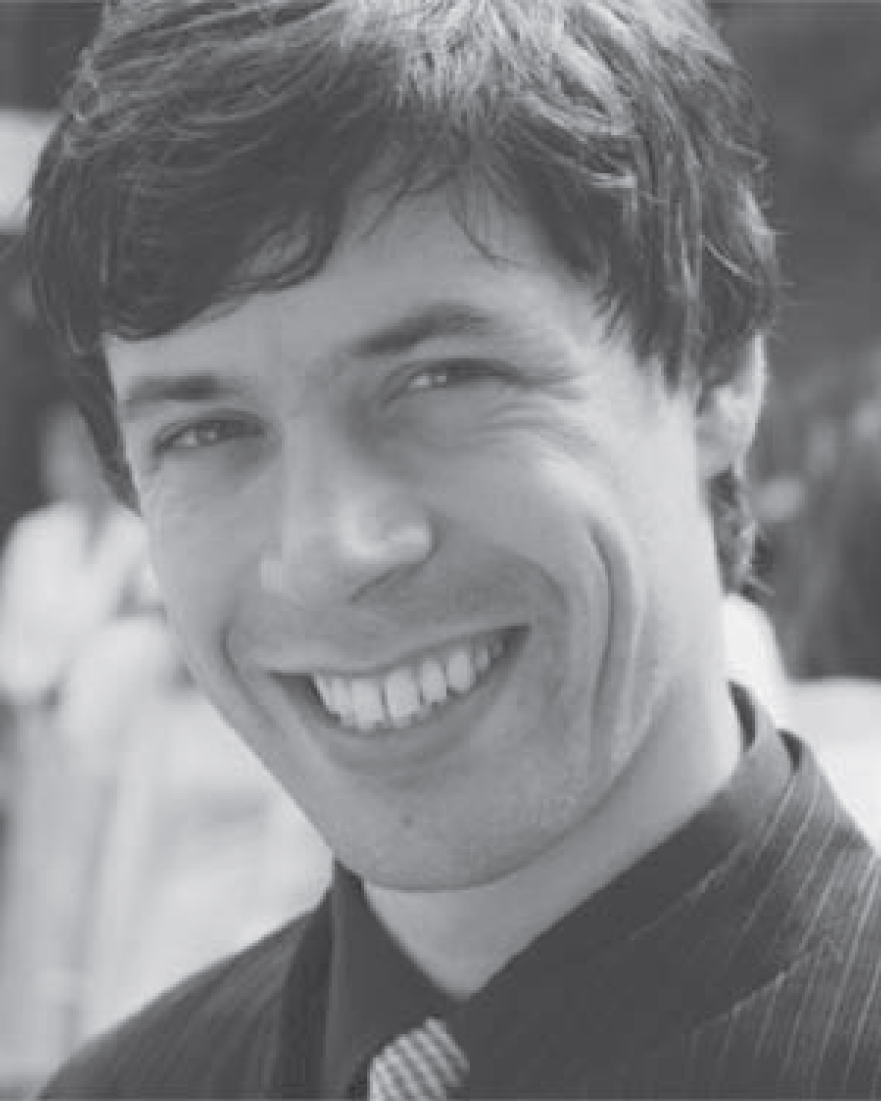}}]
{David A. Howey}
(M\textquoteright{}10) received the B.A. and M.Eng. degrees from Cambridge University, Cambridge, U.K., in 2002 and the Ph.D. degree from Imperial College London, London, U.K., in 2010. He is currently a Lecturer with the Energy and Power Group, Department of Engineering Science, University of Oxford, Oxford, U.K. He leads projects on fast electrochemical modeling, model-based battery management systems, battery thermal management, and motor degradation. His research interests include condition monitoring and management of electric-vehicle electric-vehicle components.
\end{IEEEbiography}

\vfill

\end{document}